\documentclass[aps,prl,twocolumn, groupedaddress,showpacs]{revtex4}

\usepackage{epsfig,setspace,graphics,amsmath,amssymb,bm}

\begin{document}
\title{Comment on ``Coherent Control of a V-Type Three-Level System in a Single Quantum Dot''}
\author{Yujun Zheng}
\email{yzheng@sdu.edu.cn}
\affiliation{ School of Physics and Microelectronics, 
Shandong University,  Jinan 250100,  China}


\maketitle
 
Wang {\it et al.} studied the coherent control of single Quantum Dots recently \cite{wang}. 
In Ref.~\cite{wang}, the authors model the single Quantum Dots as the V-type three level
system with two orthogonal transition dipole moments $\bm{\mu}_x$ and $\bm{\mu}_y$.
Wang {\it et al.} took the theoretical calculations via the three level system Bloch equation under the RWA. The equation they used is written as
\begin{equation}
\small{
\label{eq:bloch}
  \dot{\overrightarrow{S}}(t)=M(t) \overrightarrow{S}(t) -\Gamma \overrightarrow{S} -\overrightarrow{\Lambda} ,
}
\end{equation}
where $M(t)$ is the coefficient matrix related to external laser field. The  $\Gamma$ 
and $\overrightarrow{\Lambda}$, in Ref.~\cite{wang}, are as following 
\begin{widetext}
\begin{equation}
\small{
  \Gamma= 
  \left(
   \begin{array}{cccccccc}
   \frac{1}{2}\gamma_x & \frac{1}{2}\gamma_{xy} & 0 & \delta_x & 0 & 0 & 0 & 0 \\
   \frac{1}{2}\gamma_{xy} & \frac{1}{2} \gamma_y & 0 & 0 & \delta_y & 0 & 0 & 0 \\
   0 & 0 & \frac{1}{2}(\gamma_x+\gamma_y) & 0 & 0 & -\Delta & \frac{1}{3}\gamma_{xy} & \frac{1}{3}\gamma_{xy} \\
   -\delta_x & 0 & 0 & \frac{1}{2}\gamma_x & \frac{1}{2}\gamma_{xy} & 0 & 0 & 0 \\
   0 & -\delta_y & 0 & \frac{1}{2}\gamma_{xy} & \frac{1}{2}\gamma_y & 0 & 0 & 0 \\
   0 & 0 & \Delta & 0 & 0 & \frac{1}{2}(\gamma_x+\gamma_y) & 0 & 0 \\
   0 & 0 & \frac{1}{2}\gamma_{xy} & 0 & 0 & 0 & 0 & 0 \\ 
   0 & 0 & \frac{1}{2}\gamma_{xy} & 0 & 0 & 0 & 0 & 0 
    \end{array}
  \right), 
}
\label{eq:gamma}
\end{equation}
\end{widetext}
and
\begin{equation}
\small{
  \overrightarrow{\Lambda} =\left( 0,0,\frac{2}{3}\gamma_{xy}, 0,0,0,
                           \frac{1}{3}\gamma_x, \frac{1}{3}\gamma_y \right).
}
\label{eq:lambda}
\end{equation}

Their expressions of the coefficient matrices $\Gamma$ 
and $\overrightarrow{\Lambda}$  are wrong.
In general, the expressions of the coefficient matrices $\Gamma$ 
and $\overrightarrow{\Lambda}$ for the V-type three level system are ~\cite{peng,ficek}
\begin{widetext}
\begin{equation}
\small{
  \Gamma=
  \left(
   \begin{array}{cccccccc}
   \frac{1}{2}\gamma_x & \frac{1}{2}\gamma_{xy} & 0 & \delta_x & 0 & 0 & 0 & 0 \\
   \frac{1}{2}\gamma_{xy} & \frac{1}{2} \gamma_y & 0 & 0 & \delta_y & 0 & 0 & 0 \\
   0 & 0 & \frac{1}{2}(\gamma_x+\gamma_y) & 0 & 0 & -\Delta & \frac{1}{3}\gamma_{xy} & \frac{1}{3}\gamma_{xy} \\
   -\delta_x & 0 & 0 & \frac{1}{2}\gamma_x & \frac{1}{2}\gamma_{xy} & 0 & 0 & 0 \\
   0 & -\delta_y & 0 & \frac{1}{2}\gamma_{xy} & \frac{1}{2}\gamma_y & 0 & 0 & 0 \\
   0 & 0 & \Delta & 0 & 0 & \frac{1}{2}(\gamma_x+\gamma_y) & 0 & 0 \\
   0 & 0 & \frac{3}{2}\gamma_{xy} & 0 & 0 & 0 & \frac{1}{3}(4 \gamma_x-\gamma_y) & -\frac{2}{3} (\gamma_x-\gamma_y)\ \\ 
   0 & 0 & \frac{3}{2}\gamma_{xy} & 0 & 0 & 0 & \frac{2}{3}(\gamma_x-\gamma_y) & -\frac{1}{3}(\gamma_x - 4 \gamma_y) 
    \end{array}
  \right), 
}
\label{eq:gamma1}
\end{equation}
\end{widetext}
and
\begin{equation}
\small{
  \overrightarrow{\Lambda} =\left( 0,0,\frac{2}{3}\gamma_{xy}, 0,0,0,\frac{1}{3}(2\gamma_x+\gamma_y), \frac{1}{3} ( \gamma_x+2\gamma_y) \right) ^\dag.
}
\label{eq:lambda1}
\end{equation}

For the V-type three level system with the orthogonal excited states $|x \rangle$ and $|y \rangle$,  $\gamma_{xy}=0$~\cite{ficek}. However the authors of Ref.~\cite{wang} did not mentioned this in their theoretical calculations.  Unfortunately, the authors of Ref.~\cite{wang} did not give the parameters they used in their theoretical calculations, we can not compare the numerical results.  

\begin{acknowledgments}
I thank Y. Peng for valuable comments and suggestions.
This work was supported by the National Science Foundation of China 
(grant no. 10674083).  
Partial financial supports from the Science Foundation of Shandong Province, China.
\end{acknowledgments}

\end{document}